\documentclass[a4paper,clock]{article}
\usepackage[dvips]{epsfig}
\usepackage{amssymb}
\input{epsf}
\usepackage{bm}
\usepackage{dcolumn}

\catcode`\@=11
\def\marginnote#1{}
\newcount\hour
\newcount\minute
\newtoks\amorpm
\hour=\time\divide\hour by60 \minute=\time{\multiply\hour by60
\global\advance\minute by-\hour}\edef\standardtime{{\ifnum\hour<12
\global\amorpm={am}%
        \else\global\amorpm={pm}\advance\hour by-12 \fi
        \ifnum\hour=0 \hour=12 \fi
        \number\hour:\ifnum\minute<10
0\fi\number\minute\the\amorpm}}
\edef\militarytime{\number\hour:\ifnum\minute<10 0\fi\number\minute}

\def\draftlabel#1{{\@bsphack\if@filesw {\let\thepage\relax
   \xdef\@gtempa{\write\@auxout{\string
      \newlabel{#1}{{\@currentlabel}{\thepage}}}}}\@gtempa
   \if@nobreak \ifvmode\nobreak\fi\fi\fi\@esphack}
        \gdef\@eqnlabel{#1}}
\def\@eqnlabel{}
\def\@vacuum{}
\def\draftmarginnote#1{\marginpar{\raggedright\scriptsize\tt#1}}
\def\draft{\oddsidemargin -.5truein
        \def\@oddfoot{\sl preliminary draft \hfil
        \rm\thepage\hfil\sl\today\quad\militarytime}
        \let\@evenfoot\@oddfoot \overfullrule 3pt
        \let\label=\draftlabel
        \let\marginnote=\draftmarginnote

\def\@eqnnum{(\theequation)\rlap{\kern\marginparsep\tt\@eqnlabel}%
\global\let\@eqnlabel\@vacuum}  }

\def\numberbysection{\@addtoreset{equation}{section}
        \def\theequation{\thesection.\arabic{equation}}}

\def\underline#1{\relax\ifmmode\@@underline#1\else
 $\@@underline{\hbox{#1}}$\relax\fi}

\catcode`@=12 \relax

\numberbysection

\topmargin by -\headsep

\def\br{\begin{eqnarray}}
\def\er{\end{eqnarray}}
\def\be{\begin{equation}}
\def\ee{\end{equation}}

\def\({\left(}
\def\){\right)}

\relax

\def\tp0{\Theta_{+}^{(0)}}
\def\tm0{\Theta_{-}^{(0)}}

\def\f#1#2#3 {f^{#1#2}_{#3}}

\def\win1{{\sf w_{1+\infty}}}

\def\Win1{{\sf W_{1+\infty}}}

\def\rlx{\relax\leavevmode}
\def\inbar{\vrule height1.5ex width.4pt depth0pt}
\def\IZ{\rlx\hbox{\sf Z\kern-.4em Z}}
\def\IR{\rlx\hbox{\rm I\kern-.18em R}}
\def\IC{\rlx\hbox{\,$\inbar\kern-.3em{\rm C}$}}
\def\IN{\rlx\hbox{\rm I\kern-.18em N}}
\def\IO{\rlx\hbox{\,$\inbar\kern-.3em{\rm O}$}}
\def\IP{\rlx\hbox{\rm I\kern-.18em P}}
\def\IQ{\rlx\hbox{\,$\inbar\kern-.3em{\rm Q}$}}
\def\IF{\rlx\hbox{\rm I\kern-.18em F}}
\def\IG{\rlx\hbox{\,$\inbar\kern-.3em{\rm G}$}}
\def\IH{\rlx\hbox{\rm I\kern-.18em H}}
\def\II{\rlx\hbox{\rm I\kern-.18em I}}
\def\IK{\rlx\hbox{\rm I\kern-.18em K}}
\def\IL{\rlx\hbox{\rm I\kern-.18em L}}
\def\one{\hbox{{1}\kern-.25em\hbox{l}}}
\def\0#1{\relax\ifmmode\mathaccent"7017{#1}%
B        \else\accent23#1\relax\fi}

                \def\/{\frac}

                \def\({\Big(}
                \def\){\Big)}
                \def\[{\Big[}
                \def\]{\Big]}

  \def\rlx{\relax\leavevmode}
                \def\inbar{\vrule height1.5ex width.4pt depth0pt}
                \def\IZ{\rlx\hbox{\sf Z\kern-.4em Z}}
                \def\IR{\rlx\hbox{\rm I\kern-.18em R}}
                \def\IC{\rlx\hbox{\,$\inbar\kern-.3em{\rm C}$}}
                \def\IN{\rlx\hbox{\rm I\kern-.18em N}}
                \def\IO{\rlx\hbox{\,$\inbar\kern-.3em{\rm O}$}}
                \def\IP{\rlx\hbox{\rm I\kern-.18em P}}
                \def\IQ{\rlx\hbox{\,$\inbar\kern-.3em{\rm Q}$}}
                \def\IF{\rlx\hbox{\rm I\kern-.18em F}}
                \def\IG{\rlx\hbox{\,$\inbar\kern-.3em{\rm G}$}}
                \def\IH{\rlx\hbox{\rm I\kern-.18em H}}
                \def\II{\rlx\hbox{\rm I\kern-.18em I}}
                \def\IK{\rlx\hbox{\rm I\kern-.18em K}}
                \def\IL{\rlx\hbox{\rm I\kern-.18em L}}
                \def\one{\hbox{{1}\kern-.25em\hbox{l}}}
                \def\0#1{\relax\ifmmode\mathaccent"7017{#1}%
                B        \else\accent23#1\relax\fi}
                
\def\IC{\mathbb{C}}
\def\IR{\mathbb{R}}
\def\IK{\mathbb{K}}

\def\II{{\mathbb I}}

\def\sl{\mathfrak{sl}}

\let\phi=\varphi

\def\hA{\hat{A}}

\def\wg{\widehat{\mathfrak{g}}}
\def\hsl{\widehat{sl}(2)}

\def\hA{\hat{A}}

\def\be{\begin{eqnarray}}
\def\ee{\end{eqnarray}}

\def\>{\rangle}
\def\<{\langle}

\begin{document}
\vspace{.2in}
\begin{center}
{\large\bf Nonvanishing boundary conditions and dark solitons in the NLS model}\footnote{Poster contribution to the 5th International School on Field Theory and Gravitation,
April 20 - 24, 2009, Cuiab\'a city, Brazil}
\end{center}

\vspace{1in}

\begin{center}

L.F. dos Santos, H. Blas and M. J. B. F. da Silva

\vspace{.5 cm} \small

\par \vskip .1in \noindent

Instituto de F\'isica - Universidade Federal do Mato Grosso\\
Universidade Federal de Mato Grosso\\
Av. Fernando Correa, s/n, Coxip\'o \\
78060-900, Cuiab\'a - MT - Brazil\\
\end{center}

\vspace{2 cm}

\begin{abstract}

\vskip .1in
 We consider non-vanishing boundary conditions (NVBC) for the NLS model \cite{6,7,27} in the context of the hybrid dressing
transformation and $\tau$-function approach. In order to write the NLS model in a suitable form to deal with non-vanishing boundary conditions it is
introduced a new spectral parameter in such a way that the usual NLS parameter will depend on the affine parameter through the so-called Zukowsky function. In the context of the dressing transformation the introduction of the affine parameter avoids the construction of certain Riemann sheets for the usual NLS spectral parameter. In this way one introduces a Lax pair defined for the new spectral parameter and the relevant NVBC NLS $\tau$ functions are obtained by the dressing transformation method. We construct the one and two dark-soliton solutions explicitly.
\end{abstract}





\par \vskip .3in \noindent

\newpage
The nonlinear Schrodinger model (NLS) with vanishing or non-vanishing boundary conditions is
physically significant since it appears in many applications ranging from condensed matter to string theory
(see e.g. \cite{zarembo}). The NLS model and its multifield extensions is an integrable system (see e.g. \cite{nls} and references therein). Here we provide the affine Kac-Moody algebraic formulation of the NLS model suitably written for nonvanishing boundary conditions  and the hybrid of the dressing
and Hirota methods is used to obtain dark soliton solutions of the model.

The convenient form of the NLS for dealing with non-vanishing boundary conditions, which  support dark-soliton like solutions \cite{dark1}, can be written as
\begin{eqnarray}
\partial_t \psi + \partial_{xx} \psi - 2(|\psi|^2- \rho^2 ) \psi&=& 0.\label{ds}
\end{eqnarray}
This form of NLS model is supplied  with the non-vanishing boundary conditions given by \cite{6,7,27}
\begin{eqnarray}
\psi= \left\{\begin{array}{cl} \rho, & x \rightarrow-\infty,  \\ \rho\epsilon^2, & x  \rightarrow +\infty \end{array} \right.\hspace{2cm}\rho=\mbox{real const.},
\,\,\,\,\epsilon = e^{i\theta}.
\end{eqnarray}

In order to give a group theoretical construction of the system above, let us consider the Lax pair $A$ and $B$
\begin{eqnarray}
A&=& H^1 + \Psi^+ E_+^0 + \Psi^- E_-^0 + \Phi_1 C,\label{lax1}\\
B&=& H^2 + \Psi^+ E_+^1 + \Psi^- E_-^1 + \partial_x \Psi^+ E_+^0 - \partial_x\Psi^- E_-^0 - 2(\Psi^+ \Psi^- - \rho^2)H^0 + \phi_2C,\label{lax2}
\end{eqnarray}
where $\Psi^\pm, \,\,\phi_1$ and $\phi_2$ are the fields of the  $\hsl$ NLS model and the potential $A$ and $B$ lie in the $\hsl$ affine Kac-Moody Lie algebra.
 The Lax pair in (\ref{lax1})-(\ref{lax2}) provided with the zero-curvature condition
\begin{eqnarray}
\partial_x B - \partial_t A - \left[ A\,,\,B\right]=0,\label{NC}
\end{eqnarray}
 furnishes the model (\ref{ds}) provided that the following transformation \,$
t\rightarrow it, x\rightarrow ix, \Psi^\pm\rightarrow\Psi^\pm\epsilon^{\mp 2},
$  and  the identification
$\psi \equiv \Psi^+ = (\Psi^-)^*, $ are made. The factor $\epsilon^{\mp 2} $ is introduced for later convenience and $*$ means complex conjugation. It was considered  $\Psi^+_0\Psi^-_0 = -\rho^2$.

The vacuum solutions to be considered are the ones of constant configuration, $\Psi^\pm_0 = \rho \epsilon ^{\mp2},\,\,\phi_1 = \phi_2 = 0;$ \,
so, the vacuum fields $A_V$ and $B_V$ from (\ref{lax1})-(\ref{lax2}) are
\begin{eqnarray}
A_V= H^1 + \rho\epsilon^{-2} E_+^0 + \rho\epsilon^2 E_-^0, &\hspace{1cm}& B_V= H^2 + \epsilon^{-2}\rho E_+^1 + \epsilon^2\rho E_-^1.
\label{PLdsV}
\end{eqnarray}

The vacuum connections can be written in the form
\begin{eqnarray}
\hat A_V = \partial_x \Psi\Psi^{-1}, &\hspace{2cm}& \hat B_V = \partial_t\Psi\Psi^{-1},\label{vaccumconections}
\end{eqnarray}
where $\Psi$ is the group element
\begin{eqnarray}
 \Psi &=& (\II + k^+ E_+ - k^-E_-)e^{x\zeta\sigma_3}e^{t\kappa \sigma_3}\label{psids},
\end{eqnarray}
with $k^\pm$ being constants, $\II$ the identity matrix and $E_\pm$,  $\sigma_3$ Pauli matrices. The connections in (\ref{vaccumconections}) are called pure gauge
solutions and are solutions of the zero-curvature condition (\ref{NC}).

Considering a $2\times2$ matrix representation for (\ref{psids}) in $\hsl$ algebra, it is possible to write certain relationships between the parameters $k^+, \,k^-, \,\zeta$, \,$\kappa$ \,and\, $\lambda$

\begin{eqnarray}
\kappa = \lambda\zeta,\,\, k^{\pm} = -2 \rho \epsilon^{\mp 2} \frac{1}{\lambda +2 \zeta},\,\,\,\,4\zeta^2 = 4\rho^2 + \lambda^2.
\label{kappazetalambda}
\end{eqnarray}
The relationships  in (\ref{kappazetalambda}) show that $\zeta$ and $\kappa$ assume two possible values in terms of $\lambda$, this requires the construction of
Riemann
sheets.

Here it is introduced  an affine parameter $\xi$ such that the functions
\begin{eqnarray}\begin{array}{clcl}
\zeta =& \frac{1}{2}\left( \xi + \frac{\rho^2}{\xi}\right), &&\lambda = \xi -  \frac{\rho^2}{\xi},\\
\kappa =& \frac{1}{2}\left( \xi^2 - \frac{\rho^4}{\xi^2}\right),&& k^\pm=-\frac{\rho}{\xi}\epsilon^{\mp 2}.
\end{array}\end{eqnarray}
become single valued in terms of $\xi$. The function $\zeta(\xi)$ above is known as the Zukowsky function. 

The appearance of an affine parameter motivates us to introduce a new spectral parameter associated with the potentials $\hat A$ and $\hat B$
\begin{eqnarray}
\hat A &=& H^1 - \rho^2H^{-1} + \Psi^+E_+^0 + \Psi^- E_-^0 + \phi_1C,\label{Axi}\\
\hat B &=& H^2 + \rho^4 H^{-2} + \Psi^+(E_+^1 - \rho^2E_+^{-1})+ \Psi^-(E_-^1 - \rho^2E_-^{-1}) \nonumber\\&&\hspace{2cm}+\hspace{0,5cm} \partial_x\Psi^+E_+^0-
\partial_x\Psi^+ E_- ^0-2\Psi^+\Psi^-H^0 + \phi_2C.\label{Bxi}
\end{eqnarray}
The potentials (\ref{Axi}) and (\ref{Bxi}) written in terms of the new spectral parameter $\xi$  describe the NLS model (\ref{ds}) when the zero-curvature condition
(\ref{NC}) is used. So, the vacuum connections corresponding to (\ref{Axi}) and (\ref{Bxi}) are given by
\begin{eqnarray}
\hat A_V &=& H^1 - \rho^2H^{-1} + \rho\epsilon^{-2} E_+^0 + \rho\epsilon^2E_-^0,\label{Axivac}\\
\hat B_V &=& H^2 + \rho^4H^{-2} + \rho\epsilon^{-2} (E_+^1 - \rho^2E_-^{-1}) +\rho\epsilon^2(E_-^1 - \rho^2E_-^{-1}) -2\rho^2H^0\label{Bxivac}.
\hspace{0,5cm}
\end{eqnarray}
Notice that these potentials are deformations of the ones in (\ref{PLdsV}).

In terms of the new spectral parameter $\xi$, (\ref{psids}) takes the form
\begin{eqnarray}
\Psi = P e^{x(H^1 + \rho^2 H^{-1})} e^{t (H^2 - \rho^4 H^{-2})},
\end{eqnarray}
where
\begin{eqnarray}
P= \II - \rho\epsilon^{-2} E_+^{-1} +\rho\epsilon^{2} E_-^{-1}, &\hspace{2cm}& P^{-1}=\frac{1}{1 + \frac{\rho^2}{\xi^2}}\left( \II + \epsilon^{-2}\rho E_+^{-1} -
\epsilon^{2}\rho E_-^{-1} \right).\label{PP-1}
\end{eqnarray}
When $\rho\rightarrow 0$ one has $\xi \rightarrow \lambda$, which implies that $P \rightarrow  \II$.

\section{The dressing transformation}

The dressing transformations are non-local gauge transformations that act on the fields of the model preserving their gradation structure; they are made
with the aid of two group elements $\Theta_{+}$ and $\Theta_{-}$, such that
\begin{eqnarray}
\hA &\rightarrow& \hA^h\equiv  \Theta_\pm \hA\Theta_\pm^{-1} + \partial_x \Theta_\pm\Theta_\pm^{-1},\nonumber\\
\hat B &\rightarrow& \hat B^h\equiv  \Theta_\pm \hat B\Theta_\pm^{-1} + \partial_t \Theta_\pm\Theta_\pm^{-1}.\label{1.1}
\end{eqnarray}
It is assumed the generalized Gauss decomposition
\begin{eqnarray}
\Psi h\Psi^{-1} &=& \left(\Psi h\Psi^{-1}\right)_-\left(\Psi h\Psi^{-1}\right)_0\left(\Psi h\Psi^{-1}\right)_+\equiv\Theta_-^{-1}\hat M^{-1}\hat N.
\end{eqnarray}
The vector tau function $\vec{\tau}(x,t)$ is defined by \cite{agostinho}
\begin{eqnarray}
\vec{\tau}(x,t) &=& \left(\Psi h\Psi^{-1}\right)|\hat\lambda_0\>= \Theta_-^{-1}\hat M^{-1}|\hat\lambda_0\>\label{v-1M-1}.
\end{eqnarray}
Once the highest weight state $|\hat\lambda_0\>$ is an eigenstate of $\wg_0$ subalgebra, it is possible to define
\begin{eqnarray}
\vec{\tau}_0(x,t)=\hat M^{-1}|\hat\lambda_0\>=|\hat\lambda_0\>\hat\tau_0(x,t),
\end{eqnarray}
where $\hat\tau_0(x,t)$ is a function described by
\begin{eqnarray}
\hat\tau_0(x,t) &=& \langle\hat{\lambda}_0 |\left(\Psi h\Psi^{-1}\right)_0|\hat\lambda_0\>\label{deftau0ds}.
\end{eqnarray}

Using (\ref{v-1M-1}) and (\ref{deftau0ds}) one finds
\begin{eqnarray}
\Theta^{-1}_- |\hat\lambda_0\> &=& \frac{\vec{\tau}(x,t) }{\hat\tau_0(x,t)}.\label{Theta-1}
\end{eqnarray}

Replacing the fields  $A_V$ and $B_V$, in the form given in (\ref{PLdsV}), into the dressing transformation (\ref{1.1}), one gets
\begin{eqnarray}
\hat A^h&=& \Theta_-\left( H^1 - \rho^2H^{-1} + \rho\epsilon^{-2} E_+^0 + \rho\epsilon^2E_-^0\right)\Theta_-^{-1}+ \partial_x \Theta_-\Theta_-^{-1}\\
\hat B^h&=& \Theta_-\left(H^2 + \rho^4H^{-2} + \rho\epsilon^{-2} (E_+^1 - \rho^2E_-^{-1}) \right.\nonumber\\&& + \left.\rho\epsilon^2(E_-^1 - \rho^2E_-^{-1}) -
2\rho^2H^0\right)\Theta_- + \partial_t \Theta_-\Theta_-^{-1}
\end{eqnarray}
where $\Theta_- = \exp\left(\sum_{n>0}\hat\sigma_ {-n}\right), \,\,\hat M = \exp(\hat \sigma_0)$ and $\hat N = \exp \left(\sum_{n>0}\hat\sigma_ {n}\right).$ It is
possible
to find some of the components, say  $\hat \sigma_{-1}, \,\,\,\hat \sigma_{-2}$, in terms of the fields $\Psi^\pm, \,\,\,\phi_1$ and $\phi_2$
\be
\hat \sigma_{-1} &=& -(\Psi^+ - \rho\epsilon^{-2})E_+^{-1} + (\Psi^- - \rho\epsilon^{2})E_-^{-1} + \sigma_{-1}^0H^{-1},\\
\hat \sigma_{-2} &=& -\sigma_{-2}^+E_+^{-1} + (\Psi^- - \rho\epsilon^{2})E_-^{-1} + \sigma_{-1}^0H^{-1},
\ee
one finds $\partial_x\sigma_{-1}^0 = 2(\Psi^+\Psi^- - \rho^2), \,\,\, \sigma_{-2}^\pm = - \partial_x\Psi^\pm + \frac{1}{2}\sigma_{-1}^0\Psi^\pm, \,\,\,\,\phi_1 = -
\frac{1}{2}\sigma_{-1}^0$ and $\phi_2= -\sigma_{-2}^0$. So, with the aid of (\ref{Theta-1}) the solutions in the orbit of the vacuum are given by
\begin{eqnarray}
\Psi^+= \rho \epsilon^{-2}+\frac{\hat\tau^+}{\hat\tau^0}, &\hspace{2cm}& \Psi^-=  \rho \epsilon^{2}-\frac{\hat\tau^-}{\hat\tau^0}\label{solPsids},
\end{eqnarray}
where the $\tau^\pm$ functions are defined by
\begin{eqnarray}
\hat\tau^+ &\equiv&\<\hat\lambda_0|E_-^1\left(\Psi h\Psi^{-1}\right)_{-1}|\hat\lambda_0\>\label{deftau+},\\ \hat\tau^- &\equiv&\<\hat\lambda_0|E_+^1
\left(\Psi h\Psi^{-1}\right)_{-1}|\hat\lambda_0\>.\label{deftau-}
\end{eqnarray}

According to the dressing method, the soliton solutions are determined by choosing convenient constant group elements $h$.

\subsection{The 1-dark soliton solution}

Let us choose the group element $h=e^F, \,\,F = \sum_{n=-\infty}^\infty\nu_1^n E_-^{-n}$; the relevant $\tau$ functions are
\begin{eqnarray}
\tau^0 &=& 1 + e^{-\phi_1} \<\hat \lambda_0|PFP^{-1} |\hat\lambda_0\>\label{eltau0dseF},\\
\tau^+ &=& e^{-\phi_1} \< \hat\lambda_0|E_-^1( PFP^{-1} )|\hat\lambda_0\>\label{tau+ds0},\\
\tau^- &=& e^{-\phi_1}\<\hat \lambda_0|E_+^1( PFP^{-1}) |\hat\lambda_0\>\label{tau-ds0};
\end{eqnarray}
so the equations (\ref{solPsids}) provided the matrix elements in (\ref{eltau0dseF}), (\ref{tau+ds0}) and (\ref{tau-ds0}), furnishes the solution
 \begin{eqnarray}
\psi^+ &=& \rho\epsilon^{- 2} +  \frac{a_1e^{-\phi_1}}{1 + a_1\nu_1^2e^{-\phi_1}\rho\epsilon^{-2} (\nu_1^2 + \rho^2)}\label{solpsi1}\\
\psi^- &=& \rho\epsilon^2 + \frac{a_1\nu_1^2e^{-\phi_1}}{\rho^2\epsilon^{-4}(1 - a_1\nu_1^2e^{-\phi_1}\rho\epsilon^{-2}(\nu_1^2 + \rho^2))},\label{solpsi11}
\end{eqnarray}
where $a_1$ is a free parameter and $\phi_1 =  \left( x(\nu + \frac{\rho}{\nu}) + t(\nu^2 - \frac{\rho^4}{\nu^2})\right)$. This is just the {\bf one dark-soliton}
solution. The relevant matrix elements can be  obtained with the aid of the one level vertex operator or the integrable highest weight representations  of the $\hsl$
 Kac-Moody algebra. Similarly, one can set $h=e^{G}$, where $G =\sum_{n=-\infty}^\infty\rho_1^n E_+^{-n}$. In this way one can get another {\bf one dark-soliton}
 solution. In order to get insight into the `dark'-soliton evolution let us plot the function $(\Psi^+\Psi^-)$ for two successive times.
\begin{figure}[htb]
  \centering
  \mbox{%
      \includegraphics[scale=0.28]{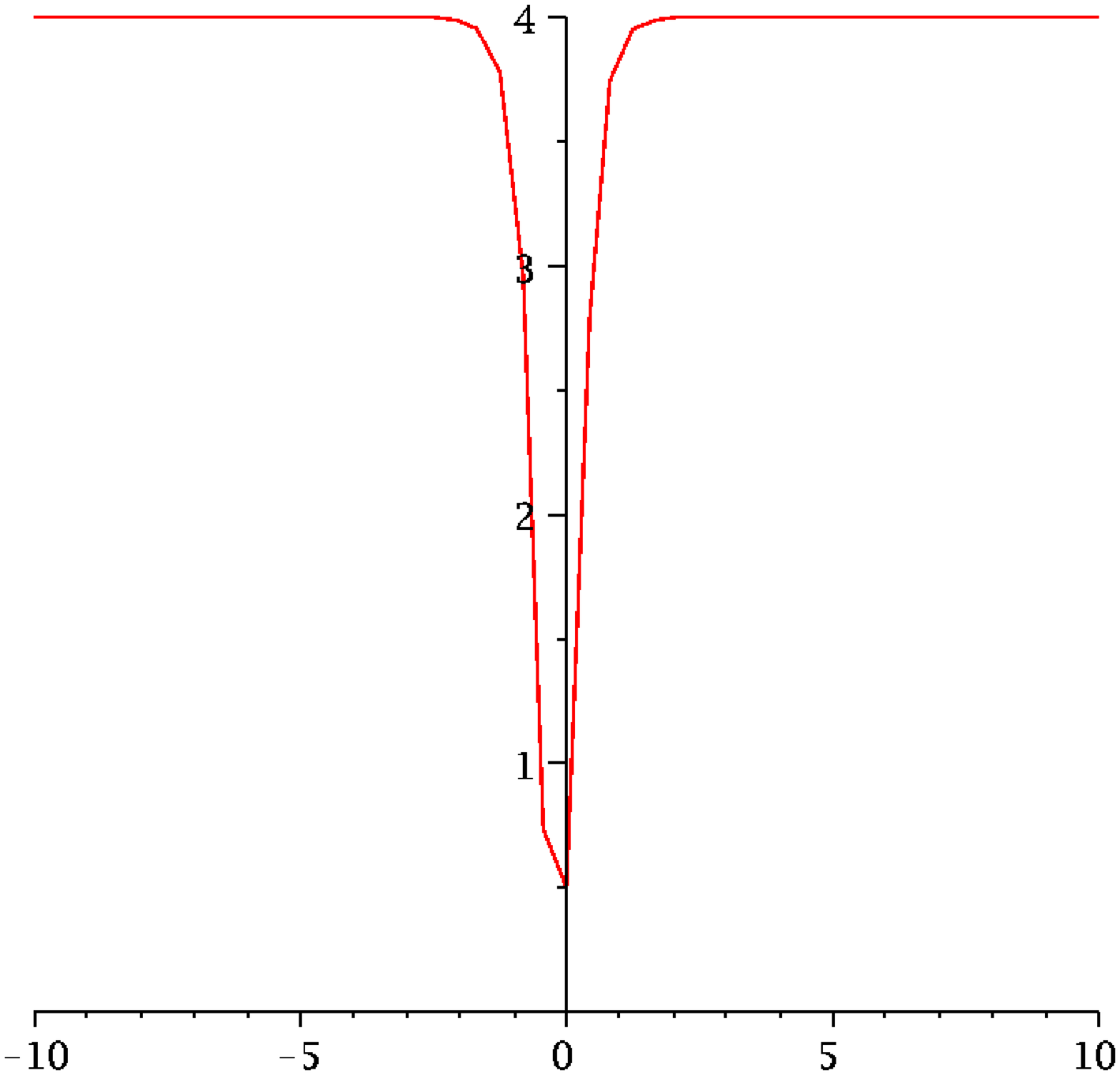}\qquad
      \includegraphics[scale=0.28]{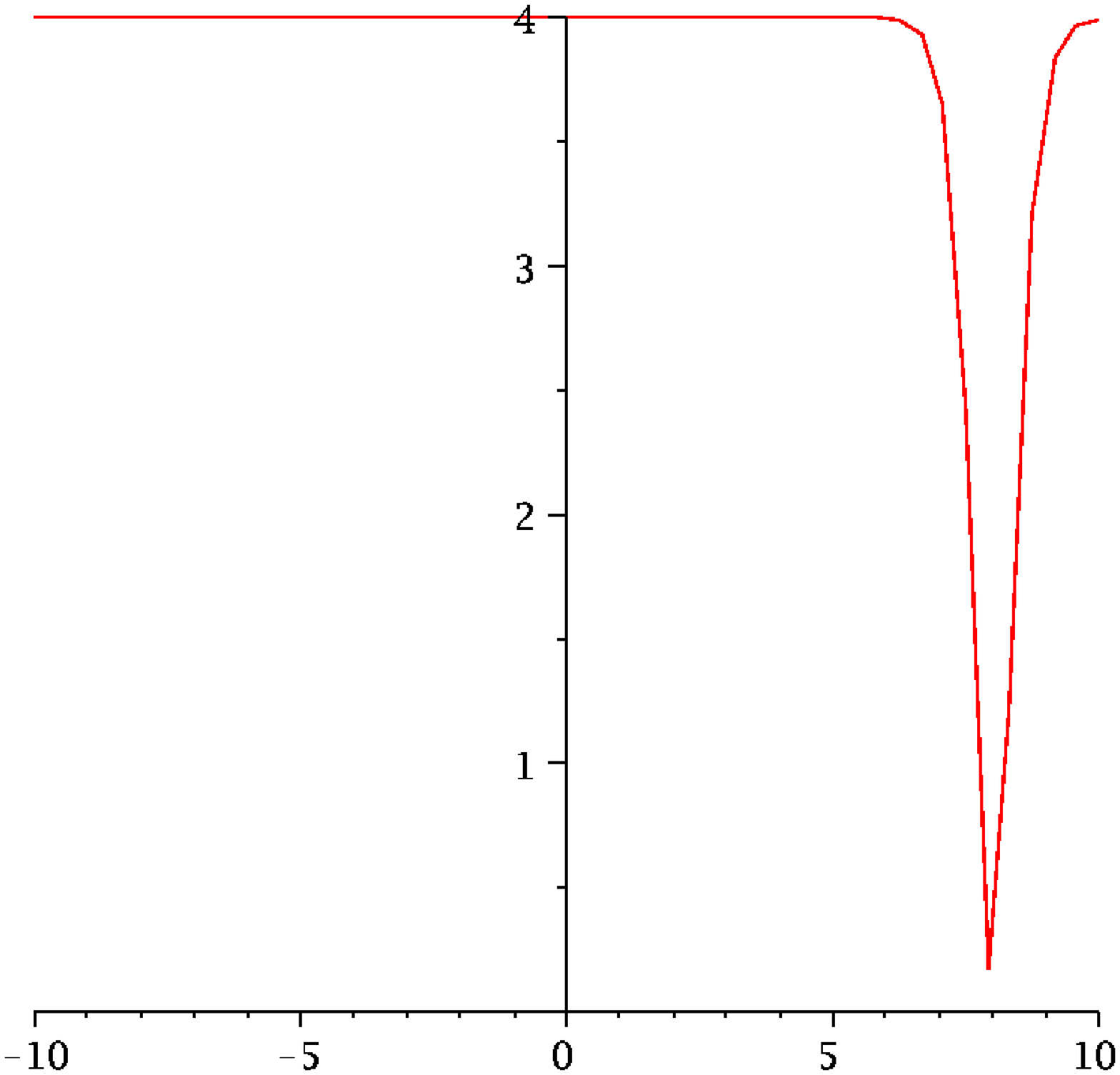}
    }\caption{1-dark soliton evolution for two successive times showing the NVBCs.}
     \label{figura1}
\end{figure}
The figure 1 is plotted for $a_1 = -2; \,\,\,\,\nu_1=1.9,\,\,\,\,\epsilon = 1, \rho=2,\,\,\,\,b_1 = -1.8,$ and $c_1 = 0.5$.  Notice the nonvanishing
boundary condition for the fields at $ x \rightarrow \pm \infty$ in the figure above.

\subsection{The 2-dark soliton solution}

In order to obtain $2-$soliton solution it is chosen $h=e^{F} e^{G}$. The relevant $\tau$ functions become
\begin{eqnarray}
\tau^0 &=& 1 + a_1e^{-\phi_1}\<\hat\lambda_0| PFP^{-1}|\lambda_0\>+ a_2e^\eta_1 \<\hat\lambda_0|PGP^{-1}|\lambda_0\> + a_3e^{\eta_1-\phi_1}
\<\hat\lambda_0|PFGP^{-1}|\lambda_0\>, \\
\tau^+ &=&  b_1e^{-\phi_1}\<\hat\lambda_0|E_-^1( PFP^{-1})|\lambda_0\>+ b_2e^{\eta_1}\<\hat\lambda_0|E_-^1(PGP^{-1})|\lambda_0\> + b_3e^{\eta_1-\phi_1}
\<\hat\lambda_0|E_-^1(PFGP^{-1})|\lambda_0\>,\hspace*{1cm} \\
\tau^- &=&  c_1 e^{-\phi_1}\<\hat\lambda_0|E_+^1( PFP^{-1})|\lambda_0\>+ c_2 e^{\eta_1} \<\hat\lambda_0|E_+^1(PGP^{-1})|\lambda_0\> + c_3e^{\eta_1-\phi_1}
\<\hat\lambda_0|E_+^1(PFGP^{-1})|\lambda_0\>.\hspace*{1cm}.
\end{eqnarray}
As usual the matrix elements above can be computed through the relevant highest weight representation of the affine Lie algebra, however one can avoid those calculations by writing these matrix elements as certain constant parameters which must be determined by direct replacement of the solutions into the relevant equations of motion. These cumbersome computations can be made with the aid of a program such as MAPLE. So, one gets the solutions $\psi^+$ and $\psi^-$ given by
\begin{eqnarray}
 \psi^+ = \rho\epsilon^{-2} + \frac{\left. a_1e^{-\phi_1} + a_2 e^{\eta_1} +
a_2a_1\frac{(m_1+ \nu_1)(\rho^2 + m_1\nu_1)^2}{\rho\epsilon^{-2}(\nu_1 - m_1)(\rho^4 + \nu_1^2 m_1^2 +
\rho^2 (\nu_1^2 + m_1^2)) }e^{-\phi_1} e^{\eta_1}\right.}
{ \left.1 - \frac{\nu_1^2a_1e^{-\phi_1}}{ \rho\epsilon^{-2} (\nu_1^2 + \rho^2) } -\frac{a_2 r_2e^{\eta_1}}{\rho^2 + m_1^2} -
\frac{(\rho^2 + \nu_1m_1)^2a_1a_2\nu_1^2e^{-\phi_1} e^{\eta_1} }{( \rho^4(\nu_1^2 - m_1^2)^2 - 2\rho^2\nu_1m_1 (m_1^2 + \nu_1^2) + \nu_1^2m_1^2(m_1 + \nu_1^2) + \rho^2(\nu_1^2 - m_1^2)^2
 )\rho^2 \epsilon^{-4} }\right.}\nonumber
\end{eqnarray}
and
\begin{eqnarray}
\psi^- = \rho\epsilon^2 +\
\frac{\left.\frac{a_1\nu_1^2}{\rho^2 \epsilon^{-4}}e^{-\phi_1}+ \frac{a_2\rho^2\epsilon^4 }{m_1^2}e^{\eta_1}
+ a_1a_2\nu_1^2\rho\epsilon^2\frac{(m_1+\nu_1)(\rho^2 + m_1\nu_1)^2}{m_1^2\rho^2\epsilon^{-4}(m_1-\nu_1)(\rho^4 + \nu_1^2 m_1^2 +\rho^2( m_1^2 +
 \nu_1^2) )}e^{-\phi_1} e^{\eta_1}\right.}
{ \left.1 - \frac{\nu_1^2a_1e^{-\phi_1}}{ \rho\epsilon^{-2} (\nu_1^2 + \rho^2) } -\frac{a_2r_2e^{\eta_1}}{\rho^2 + m_1^2} -
\frac{(\rho^2 + \nu_1m_1)^2a_1a_2\nu_1^2e^{-\phi_1} e^{\eta_1} }{( \rho^4(\nu_1^2 - m_1^2)^2 - 2\rho^2\nu_1m_1 (m_1^2 + \nu_1^2) + \nu_1^2m_1^2(m_1 + \nu_1^2) + \rho^2(\nu_1^2 - m_1^2)^2
 )\rho^2 \epsilon^{-4} }\right.}\nonumber
\end{eqnarray}
where $\phi_1=x (\nu_1+\frac{\rho^2}{\nu_1})+(\nu_1^2-\frac{\rho^4}{\nu_1^2})t, \,\,\eta_1 = x (m_1+\frac{\rho^2}{m_1})+(m_1^2-\rho^4/m_1^2) t$, and $r_{2}= \rho \epsilon^2$.
The corresponding 2-dark soliton evolution can be visualized by plotting the function $(\psi^+\psi^-)$ for certain parameter values (see Fig. 2).
\begin{figure}[htb]
  \centering
\includegraphics[scale=0.21]{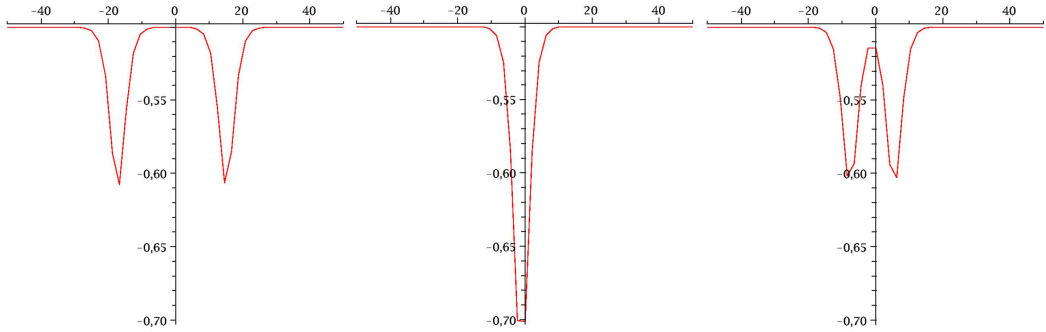}
\caption{2-dark soliton evolution for three successive times.}
    \end{figure}

{\bf Acknowledgements}

LFS and MJBS thank CAPES for financial support, and HB thanks CNPq for partial support.


\begin{thebibliography}{99}
\bibitem{6} E. V. ~Doktorov, \emph{J. Math Phys.} {\bf 38} (1997) 4138.

\bibitem{7} V. S. ~Gerdjikov, \emph{Selected aspects of soliton theory constant boundary conditions}, [{\tt nlin/0604005}].

\bibitem{27} L. D. ~Faddeev and L.A. Takhtajan, \emph{Hamiltonian Methods in the Theory of Solitons}, Springer-Verlag, London, (1987).

\bibitem{zarembo} K. ~Zarembo, \emph{Quantum Giant Magnons},
  \emph{JHEP} {\bf 08} (047) 005 [{\tt hep-th/08023681}].

\bibitem{nls}
  H.~Blas, L.~A.~Ferreira, J.~F.~Gomes and A.~H.~Zimerman,
  \emph{Phys.\ Lett.} {\bf A237} (1998) 225, [{\tt solv-int/9701012}].

\bibitem{dark1} H. ~Blas, M. J. B. F. da Silva and L. F. dos Santos, \emph{Generalized NLS bright and dark solitons in the hybrid dressing and tau function approach}, (to appear).

\bibitem{agostinho}
L. A. Ferreira, J.L. Miramontes, J.S.-Guillen \emph{J.Math.Phys.} {\bf 38} (1997) 882.\\
H.~Blas, \emph{Vector NLS hierarchy solitons revisited: dressing transformation and tau function approach},
  [{\tt solv-int/9912015}].
\end{thebibliography}
\end{document}